\documentclass{iau} 
\usepackage{graphicx}

\newcommand{\Teff}{$T_{\rm eff}$}
\newcommand{\Vt}{V$_{\rm mic}$}

\newcommand{\kms}{km\,s$^{-1}$}

\newcommand{\aap}{A\&A}

\newcommand{\mnras}{MNRAS}

\title[NLTE abundances of C, O, Ca, Ti, and Fe in the reference BAF-type stars] 
{NLTE abundances of C, O, Ca, Ti, and Fe in the reference BAF-type stars}

\author[Tatyana Sitnova et al.]   
{Tatyana Sitnova, Tatyana Ryabchikova,\\ Sofya Alexeeva, Lyudmila Mashonkina}

\affiliation{Institute of Astronomy, Russian Academy of Sciences, \\ Pyatnitskaya 48, 119017, Moscow, Russia \\ email: {\tt sitnova@inasan.ru} \\[\affilskip]}

\pubyear{2017}
\volume{334}  
\jname{Rediscovering our Galaxy}
\editors{C. Chiappini, I. Minchev, E. Starkenburg, M. Valentini, eds.}
\begin{document}

\maketitle

\begin{abstract}
We present accurate methods of abundance determination based on the non-local thermodynamic equilibrium (NLTE) line formation for carbon, oxygen, calcium, titanium, and iron in the atmospheres of BAF-type stars. For C~I-II, O~I, Ca~I-II, and Ti~I-II, their comprehensive model atoms were described in our previous papers. A fairly complete model atom of Fe~I-II is first applied in this study. 
We determine the NLTE abundances of the nine BAF-type stars with well-determined atmospheric parameters, using high-resolution and high signal-to-noise ratio spectral observations in the broad wavelength range, from the UV to the IR. For C, Ca, Ti, and Fe, NLTE leads to consistent abundances from the lines of the two ionisation stages.
The C~I, Ca~II, and Fe~II emission lines were detected in the near IR spectrum of the late B-type subgiant star HD~160762. They are well reproduced in the classical hydrostatic model atmosphere, when applying our NLTE methods.
\end{abstract}


Chemical abundances of BAF-type unevolved stars represent abundances of the galactic matter at modern epoch. 
Models of stellar interior structure and evolution depend on opacity and require accurate chemical abundances.
Iron is a key element for determination of stellar atmosphere parameters.
To derive accurate element abundance, one needs to explore how the deviations from local thermodynamic equilibrium (LTE) affect the formation of the element lines.
For the first time, we investigate the statistical equilibrium of iron in the atmospheres of BAF stars using a $comprehensive$ model atom of Fe~I-II.
The model atom presented by \cite{mash_fe} was upgraded by including the measured and predicted high-excitation ($> 10$~eV) energy levels of Fe~II, as available in R.~Kurucz database ({\tt http://kurucz.harvard.edu/atoms.html}). For BAF stars, NLTE leads to weakened Fe~I lines and to positive NLTE abundance corrections, in agreement with \cite{1996A&A...312..966R}, however, their magnitude is smaller compared to that in the earlier paper. Lines of Fe~II are strengthened in NLTE, and the effect grows with line strength.
The statistical equilibrium calculations for C~I-II, O~I, Ca~I-II, and Ti~I-II were performed using the methods described in \cite{2016MNRAS.462.1123A,2013AstL...39..126S,mash_ca}, and \cite{2016MNRAS.461.1000S}, respectively.

We selected the nine well-studied stars, which do not reveal any pulsation activity, chemical stratification, and magnetic field. Atmospheric parameters were derived in our earlier studies by the common method based on multi-colour photometry, analysis of Balmer lines and metal lines in high-resolution spectra, and a comparison of the spectrophotometric data with the theoretical flux. They cover the following ranges in effective temperature:
7250~K $\le$ \Teff\ $\le$ 17500~K, surface gravity: 3.55 $\le$ log~g $\le$ 4.30, and metallicity: $-0.30 \le$ [Fe/H] $\le$ 0.45.

We determined the NLTE and LTE abundances of carbon, oxygen, calcium, titanium, and iron in the reference stars.
For each species, NLTE leads to smaller line-to-line scatter compared to LTE.
For the majority of stars, we obtained smaller differences in the NLTE than in the LTE abundances between the neutral and singly ionised species. For example, in LTE, the abundance discrepancy between Fe~I and Fe~II reaches 0.22~dex.
The exception is HD~17081, where Fe~I and Fe~II agree in LTE, while NLTE leads to Fe~I -- Fe~II = 0.11~dex. For the same star, we obtain C~I -- C~II = 0.13~dex in NLTE, while $-0.31$~dex in LTE. An upward revision of \Teff, by about 80~K, can provide consistent NLTE abundances from neutrals and first ions in HD~17081.

Lines of Fe~I and Fe~II are widely used for atmospheric parameter determination.
We evaluated how NLTE affects the results.
For atmosphere with \Teff/log g/[Fe/H]/\Vt\  = 10000/4.0/0.0/2.0, NLTE leads to an average abundance shift  Fe I (NLTE) -- Fe I (LTE) = 0.09~dex.
To get consistent abundances from Fe~I and Fe~II in LTE, one needs to assume either 0.20~dex lower log~g or 130~K higher \Teff, when the other parameter is fixed.
NLTE leads to a downward revision of the microturbulent velocity because the deviations from LTE are larger for stronger lines of Fe~II. For example, LTE requires an 0.2\,\kms\ larger \Vt\ for HD~145788 (9750/3.70/0.0/1.8) than in NLTE.
We found a qualitatively similar, but stronger effect on the \Vt\ determinations based on the Ti~II lines (see \cite[{Sitnova} {et~al.} 2016]{2016MNRAS.461.1000S} for details). 

In the near IR spectrum of HD~160762 (17500/3.8), the hottest star of our sample, we found emission lines of C~I, Ca~II, and Fe~II. These lines have the same Doppler width and radial velocity as the absorption spectral lines in HD~160762. We explain these phenomena by the NLTE effects in the classical plane-parallel model atmosphere.
The C~I emission line formation was described in details by \cite{2016MNRAS.462.1123A}. 
In the atmosphere of HD~160762, Fe~III is the majority species, such that N(FeIII)/N(FeII) $\ge$ 100.  Overionisation of Fe~II results in depleted energy levels and, hence, weakened Fe~II lines. In case of the high-excitation lines, they either disappear or come into emission. The same happens with the  Ca~II high-excitation lines.

{\underline{\it Conclusions}}.
The NLTE calculations with a comprehensive model atom of Fe~I--II were performed for the first time for the selected BAF-type stars. For the same  stars, we determined the 
NLTE and LTE abundances of carbon, oxygen, calcium, titanium, and iron.
For each element, NLTE reduces the line-to-line scatter compared to the LTE case and leads to consistent abundances from lines of the neutral and singly-ionised species.  
We can reproduce the emission lines of C~I, Ca~II, and Fe~II observed in the near-IR spectrum of the late B-type star HD~160762 using our NLTE methods with the classical plane-parallel and LTE model atmosphere. 
For the atmospheric parameters  of the A0 main-sequence star,
the LTE analysis of Fe~I and Fe~II lines leads to an overestimation of \Teff, by 130~K, or an underestimation of log~g, by 0.20~dex. 

L.M. and T.S. thank a support from the International Astronomical Union and the Russian Foundation for Basic Research (15-02-06046) of the participation at IAUS334.
We made use the SIMBAD, MARCS, and VALD databases.

\end{document}